\documentclass[twocolumn,tighten,time]{aastex631}
\usepackage[fleqn]{amsmath}
\usepackage{graphicx}	
\pdfoutput=1 
\usepackage{epsfig}
\usepackage{subfigure}
\usepackage{multirow}
\usepackage{hyperref}

\newcommand{\der}{{\rm d}}

 \newcommand{\m}{_{\rm m}}  
  
 \newcommand{\cc}{_{\rm c}}

\newcommand{\pk}{_{\rm pk}}

\newcommand{\ta}{_{\rm ta}}

\newcommand{\F}{^{\rm th}}  
 \newcommand{\nb}{_{\rm a}}

\newcommand{\maxi}{_{\rm max}} 
\newcommand{\modot}{M$_\odot$\ } \newcommand{\modotc}{M$_\odot$}
\newcommand{\ti}{t_{\rm i}}

\newcommand{\beq}{\begin{equation}} \newcommand{\eeq}{\end{equation}}
 \newcommand{\beqa}{\begin{eqnarray}}
\newcommand{\eeqa}{\end{eqnarray}} \newcommand{\lav}{\langle}
\newcommand{\rav}{\rangle} 
\newcommand{\vir}{_{\rm vir}}

\newcommand{\no}{_{{\rm a}1}} \newcommand{\nd}{_{{\rm a}2}} \newcommand{\nt}{_{{\rm a}3}}
 \newcommand{\tc}{t_{\rm c}} 

\begin{document}

\shorttitle{Protohalo Angular Momemtum}
\shortauthors{Salvador-Sol\'e \& Manrique}

\title{THE TIDAL TORQUE THEORY REVISITED. I. PROTOHALO ANGULAR MOMENTUM}

\author{Eduard Salvador-Sol\'e and Alberto Manrique}
\affiliation{Institut de Ci\`encies del Cosmos. Universitat de Barcelona, E-08028 Barcelona, Spain}

\email{e.salvador@ub.edu}



\begin{abstract}
In the tidal-torque theory, the angular momentum (AM) of dark matter halos arises from the tidal torque suffered by aspherical collapsing patches due to surrounding mass fluctuations. This theory was implemented in the peak model where protohalos are ellipsoidal. However, the adopted delimitation of these objects was doubtful and the protohalo AM was calculated numerically, which complicated the interpretation of the result and compromised its applicability. In addition, the AM of final halos was derived without taking into account non-linear effects. Here, we re-derive the protohalo AM in the peak model, delimiting ellipsoids in the usual natural way and following a novel fully analytic approach that leads to a very simple and practical expression. The predicted AM is shown to fully agree with the results of cosmological simulations. In Paper II, we will apply this model to infer the rotational properties of relaxed halos, accounting for shell-crossing and major mergers. 
\end{abstract}




\section{INTRODUCTION}\label{intro}

Dark matter halos are believed to acquire their angular momentum (AM) through the tidal torque of neighboring mass fluctuations on their seeds \citep{Hea49}. Although \citet{P69} found that this mechanism, the so-called tidal-torque theory (TTT), does not work for spherical protohalos, \citet{D70} and \citet{W84} showed that it does in case of other shapes due to the misalignment of the protohalo inertia tensor with respect to the gravitational tidal tensor. 

Specifically, in the \citet{Z70} approximation holding to first order in perturbed quantities, by Taylor expanding to second order the deformation tensor around the center of mass (c.o.m.) of the protohalo and integrating over its volume, \citet{W84} found that the Cartesian components of the protohalo AM, {\bf J}, grow with time $t$ according to
\beq
J_i(t) \approx - a^2(t)\dot D(t)\,\epsilon_{ijk}{T}_{jl}{I}_{lk},
\label{1}
\eeq
where $\epsilon_{ijk}$ is the fully antisymmetric Levi-Civita rank-three tensor, ${T}_{jl}=\partial^2 \phi/\partial x_j\partial x_l$ is the tidal tensor (equal to the shear or deformation tensor in the linear regime), i.e.~the Hessian of the {\it linear} gravitational potential $\phi$ at the protohalo c.o.m., and $I_{lk}$ is the protohalo inertia tensor with respect to that point. The growth of $J_i$ is thus encoded in the factor $a^2(t)\dot D(t)$, where $a(t)$ and $D(t)$ are the cosmic scale and linear growth factor, respectively, and a dot denotes time-derivative, the remaining factor being the so-called Lagrangian protohalo AM, independent of the arbitrary initial time $\ti$ where it is calculated. 

The AM growth of protohalos was studied analytically \citep{W84,H86} and numerically (\citealt{EJ79,BE87,Cea96b,Sea00}, hereafter SSK00; \citealt{Pea02a,Pea02b}. The results confirmed the validity of Equation (1) roughly until protohalos reach turnaround and rapidly contract so that the AM basically freezes out \citep{P69}. In addition, the AM of halos of mass $M$ was found to be proportional to $M^{5/3}$ (but see \citealt{Lea15}). 

To determine the {\it typical} AM of halos and understand the origin of the $J\propto M^{5/3}$ relation, \citet{HP88}, \citet{H88}, and \citet{Cea96a} (see also \citealt{R88,QB92,EL95}) implemented Equation (1) in the peak model of structure formation, where collapsing patches, marked by triaxial density maxima in the Gaussian-smoothed density field, are naturally ellipsoidal \citep{D70}. However, at that moment, the collapse time and extension (mass) of protohalos associated with peaks were unknown, so these authors adopted the top-hat spherical collapse and delimited ellipsoids in an unusual arguable way. In addition, the average over the sharpness, shape, and shear field of peaks prevented them from obtaining a practical final expression. 

However, since the publication of these works, a new formalism has been developed (see the recapitulation in \citealt{SM19}) that provides the tools appropriate to address these issues. Indeed, the so-called {\it ConflUent System of Peak Trajectories} (CUSP) has already allowed one to derive, directly from peak statistics and with no free parameters, the halo density and kinematic profiles \citep{Sea12a,Sea12b,Sea3}, substructure \citep{I,II,III}, mass function \citep{Jea14b}, and primary and secondary biases \citep{SM24,Sea24}, leading in all cases to theoretical predictions in full agreement with the results of simulations. 

The aim of the present Paper and a forthcoming one (Paper II) is to apply the TTT to derive the typical halo AM in the peak model. Here, we revisit the application of this theory to protohalos in the linear regime, following a novel approach that remedies the shortcomings of previous works. Instead of dealing with the global shear field at the location of the protohalo, we split it into the torque of individual neighboring (positive and negative) mass fluctuations. This allows us to clarify the origin of the typical protohalo AM properties and to make physically motivated approximations leading to a simple final expression that can be readily checked against simulations. In this strategy, CUSP plays a crucial role as it provides: i) accurate masses and collapse times of ellipsoidal patches \citep{Jea14a}; ii) the connection between neighboring mass fluctuations of different scales \citep{MSS95}; and iii) the correlation between all these objects \citep{SM24}. In Paper II this model will be used to derive the rotational properties of halos, taking into account the effects of shell-crossing and major mergers thanks to the results reported in \cite{Sea12a} and \citet{SM19}.

The layout of the Paper is as follows. In Section \ref{CUSP} we recall some elements of CUSP used in the derivation. In Section \ref{stra}, we describe our strategy and the planning of the work. The protohalo inertia tensor and the tidal tensor of neighboring mass fluctuations are derived in Sections \ref{correspondence} and \ref{potential}, respectively. In Section \ref{AM} we compute the typical protohalo AM. Our results are summarized and discussed in Section \ref{sum}.

\section{The CUSP formalism}\label{CUSP}

Next, we briefly explain some results of CUSP that are used in our derivation. Interested readers are referred to the cited works for more details. 

\subsection{Accurate Protohalo Extension and Collapse Time}

As well known, the collapse time $\tc$ of ellipsoidal patches at $\ti$ depends not only on their size and mass, like in spherical collapse, but also on their concentration and shape (e.g.~\citealt{P80}), or equivalently, on the scale $R$, height $\nu$ (the density contrast $\delta$ scaled to its r.m.s.~value $\sigma_0$), curvature or sharpness $x$ (minus the Laplacian scaled to its r.m.s.~value $\sigma_2$), ellipticity $e$, and prolateness $p$ of the associated peaks in the Gaussian-smoothed density field. However, $e$ and $p$ depend only on $x$ (see below), whose probability distribution function (PDF) is very sharply peaked, so all peaks with fixed $\delta$ and $R$ collapse essentially at the same time $\tc$. 

In these circumstances, choosing the scale $R(M,\tc,\ti)$ of protohalos with fixed $\delta$ at $\ti$ that evolve at $\tc$ (in the cosmology under consideration) into haloes of different masses $M$ (according to the chosen halo mass definition, i.e.~their characteristic overdensity), all protohalos associated with peaks with $\delta(\tc,\ti)$ at scale $R(M,\tc,\ti)$ will collapse, by construction, at the same time, regardless of their mass $M$. 

\citet{Jea14a} showed that the functions $M(R,\tc,\ti)$ and $\tc(\delta,\ti)$ setting the mass and collapse time of protohalos associated with Gaussian peaks with $\delta$ at $R$ are fully determined by the consistency conditions that: i) the halo mass function predicted by CUSP is correctly normalized, and ii) the halo density profile predicted by CUSP leads to the mass used to derive it. Specifically, writing the density contrast $\delta$ and scale $R$ of peaks at $\ti$ as proportional to their well-known counterpart in top-hat spherical collapse, denoted by index `th', 
\beq
\delta(\tc,\ti)=r_\delta(\tc)\delta\F(\tc,\ti)=r_\delta(\tc)\delta\cc\F(\tc)\frac{D(\ti)}{D(\tc)}~~
\label{deltat}
\eeq
\vspace{-15pt}
\beqa
\sigma_0(R,\ti)=r_\sigma\!(M,\tc)\sigma_0\F(M,\ti)~~~~~~~~~~~~~~~~~~~~~\nonumber\\
=r_\sigma(M,\tc)\sigma_0\F(M,\tc)\frac{D(\ti)}{D(\tc)},~~~~~~~~~~~
\label{rm0}
\eeqa
where $\delta\F\cc(t)$ is the critical linearly extrapolated density contrast for top-hat spherical collapse at $t$ (equal to 1.686 in the Einstein-de Sitter (EdS) cosmology), and the proportionality functions $r_\delta(\tc)$ and $r_\sigma(M,\tc)$ are well fitted by specific analytic functions dependent on cosmology and halo mass definition given in \citet{SM24}, though approximately satisfying $r_\delta(\tc)\approx a(\tc)/D(\tc)$ and $r_\sigma(M,\tc)\approx 1$ in all cases.

These relations imply in turn
\beq
\nu(M,\tc)=\frac{r_\delta(\tc)}{r_\sigma(M,\tc)}\nu\F(M) ~\approx \frac{a(\tc)}{D(\tc)}\nu\F(M),
\label{nus}
\eeq
showing that the Gaussian height of peaks associated to protohalos collapsing at $\tc$ {\it is directly related to its mass $M$}, as in top-hat smoothing. Note that this relation is independent of $\ti$.

Equation (\ref{rm0}) implicitly gives the scale $R$ as a function of $M$, $\tc$, and $\ti$ through the 0th Gaussian and top-hat spectral moments. However, for power-law spectra $P(k,t_0)=P_0 k^{n}$, where $t_0$ is the present time, $R$ can be directly related to its top-hat counterpart $R\F$, satisfying, at leading order, $M=(4\pi/3) \bar\rho(\ti) (R\F)^3$, where $\bar\rho(t)$ is the mean cosmic density at $t$. Indeed, in this case, the $j$th spectral moments for Gaussian and top-hat filters $f$ read
\beq
(\sigma_j^f)^2(R^f,t)\!=\!\frac{P_0 [D_0/D(t)]^2}{2\pi^2 (R^f)^{n+3+2{\rm j}}}\!\!\int_0^\infty\!\! \der x\, x^{n+2(1+{\rm j})}\, (W^f)^2(x),
\label{sigmas}
\eeq
where $D_0=D(t_0)$, so their ratio leads to 
\beq
R(M,\tc,\ti)=r_{\rm R}(M,\tc)\,R\F(M,\ti),
\label{rmt}
\eeq
with
\beq
r_{\rm R}(M,\tc)\equiv \left[K_n\, r_\sigma(M,\tc)\right]^{1/m}\!,
\label{rm}
\eeq
where $K^2_n\equiv\int_0^\infty \der x \, x^{n+2}\,(W\F)^2(x)/\int_0^\infty \der x \, x^{n+2}\,W^2(x)$, being $W$ and $W\F$ the Fourier transforms of the Gaussian and top-hat filters, respectively, and $m\equiv -(n+3)/2$. Since the cold dark matter (CDM) spectrum is locally a power-law, for galaxy mass halos we can also adopt the relation (\ref{rm}), with $n\approx -1.75$, $m\approx -0.63$, and $K_n\approx 1.6$. This leads to $r_{\rm R}$, a function of $M$ (explicitly and through index $n$) and $t\cc$, except for virial masses (i.e. masses encompassing virial overdensities (\citealt{bn98}) with respect to the mean cosmic density), in which case $r_\sigma$ depends only on $M$ \citep{SM24}, so $r_R$ does, too. 

It may be argued that these protohalo masses and collapse times rely on CUSP. That is true, but there is no doubt about the goodness of CUSP, as evidenced by all its many successful results, listed in Section \ref{intro}. 

For simplicity in the notation, we skip from now on, unless necessary, the argument $\ti$ in all quantities referring to the (arbitrary) initial time, as well as the argument $R$ of the spectral moments $\sigma_j$.

\subsection{Continuous Peak Trajectories}\label{traj}

The functions $\delta(\tc)$ and $R(M,\tc)$ define a correspondence between halos and peaks. 
Indeed, the general relation
\begin{equation}
\frac{\partial\delta({\bf r},R)}{\partial R}=R\nabla^{2}\delta({\bf r},R)\equiv-x({\bf r},R)
R \sigma_{2}
\label{Gau}
\end{equation}
holding for Gaussian smoothing shows that, when the smoothing scale $R$ is increased, the density contrast $\delta$ of individual peaks decreases, which conforms with the fact that the functions $\delta(t)$ and $R(M,t)$ in that correspondence must be monotonically deceasing and increasing functions of $t$ and $M$, respectively. Moreover, the relation (\ref{Gau}) allows one to identify peaks (essentially at the same fixed point) tracing the same halo at infinitesimally close smoothing scales $R$ \citep{MSS95}. 

Therefore, the mass growth of any individual accreting halo traces a continuous peak trajectory in the $\delta$-$R$ plane, solution for the suited boundary condition of the differential equation (see Equation (\ref{Gau})), 
\begin{equation}
\frac{\der \delta}{\der R}=-x[R,\delta(R)]R\sigma_{2},
\label{eq}
\end{equation}
where $x[R,\delta(R)]$ is the curvature of the peak at the point $(\delta(R),R)$ of the trajectory. Individual peak trajectories are hard to calculate because they depend on the particular realization of the density field around the peak. However, they zigzag around the {\it mean} peak trajectory, solution of Equation (\ref{eq}) with the curvature $x[R,\delta(R)]$ replaced by the mean curvature $\lav x\rav[R,\delta(R)]$ and the same boundary condition. We can thus adopt such mean trajectories for the typical evolution of peaks in the $\delta$-$R$ plane. 

The mean curvature of peaks with $\delta$ at $R$ is (BBKS)  
\beq
\lav x\rav(R,\nu)=\frac{G_1(\gamma,\gamma\nu)}{G_0(\gamma,\gamma\nu)},
\eeq 
where
\beqa
G_i(\gamma,\gamma\nu)\!=\!\int_0^\infty\! \der x\,x^i\,F(x) \frac{{\rm e}^{-\frac{(x-\gamma\nu)^2}{2(1\!-\!\gamma^2)}}}{[2\pi(1-\gamma^2)]^{1/2}}~~~~~~~~~\label{G}\\
F(x)\!\equiv \!\frac{(x^3\!-\!3x)\left\{{\rm erf}\left[\!\left(\frac{5}{2}\right)^{\frac{1}{2}} x\!\right]\!+\!{\rm erf}\left[\!\left(\frac{5}{2}\right)^{\frac{1}{2}}\frac{x}{2}\!\right]\right\}}{2\!+\!\left(\frac{2}{5\pi}\right)^{\frac{1}{2}}\!\!\left[\!\left(\frac{31x^2}{4}\!+\!\frac{8}{5}\right){\rm e}^{-\frac{5x^2}{8}}\!+\!\left(\frac{x^2}{2}\!-\!\frac{8}{5}\right)
{\rm e}^{-\frac{5x^2}{2}}\!\right]}\!,\nonumber
\eeqa
$\gamma\equiv \sigma^{2}_{1}/(\sigma_{0}\sigma_{2})$. As shown by BBKS, $\lav x\rav$ takes the form $\gamma\nu+\theta(\gamma,\gamma\nu)$ with the function $\theta$ negligible at large scales as corresponding to the large-scale mass fluctuations causing torques (see Sec.~\ref{potential}). Consequently, in the case of power-law spectra (and the CDM spectrum) for which $R^2\sigma_2 \propto\sigma_0$, the mean trajectories of such peaks satisfy (see Equation (\ref{eq}))
\beq
\frac{\der \ln\delta}{\der \ln R}=m.
\label{eqtraj}
\eeq
For simplicity in the notation, we skip from now on, unless necessary, the arguments of the curvature moments  and write $\lav x\rav$, $\lav x^2\rav$, and so on. 

Continuous peak trajectories are interrupted when the corresponding accreting halos merge. Thus, trajectories starting at peaks with $\delta_0$ at $R_0$ will reach a typical maximum scale $R\maxi(R_0,\delta_0)$ equal to the mean separation between peaks at that scale, when the corresponding accreting halos typically come in contact and merge.

\subsection{Correlation between Peaks}\label{corr}
 
At first order of the `perturbative bias expansion', the correlation $\xi_{\rm p}(r)$ between peaks with $\delta$ at $R$ reads $\xi_{\rm p}(r)=B_1^2(R,\nu)\xi(r)$, where $\xi(r)$ is the matter correlation function (well fitted by $(r/s_0)^{-\tilde\gamma}$, with $\tilde\gamma\approx 2$ and $s_0\sim 15 h^{-1}$ Mpc; \citealt{Aea24}) and $B_1(R,\nu)$ is the Lagrangian linear peak bias, given by  \citep{SM24}
\beq
B_1(R,\nu)= q\,\frac{\nu-\gamma\,\lav x^2 \rav/\lav x \rav}{\sigma_0\,(1-\gamma^2)},
\label{simple1}
\eeq
with $q$ equal to $2^{(n+3)/2}$ for power-law power spectra in general and about $1.6$ in the case of the CDM spectrum.

\subsection{Peaks and Holes}\label{holes}

The peak model was developed to deal with protohalos as local maxima in the linear Gaussian random density field at $\ti$. But in this Paper we will also be concerned with local minima or holes. Fortunately, the statistics of peaks is the same as of holes, except for the sign of the eigenvalues of the Laplacian of the density field at the peak. Thus, all expressions derived for peaks can be readily extended to holes simply by changing the sign of the trace $x$ of the scaled Laplacian. (The ellipticity and prolateness of peaks and holes are also defined in terms of the Laplacian eigenvalues, though they are always defined with positive sign.) 

This comment applies, in particular, to the continuous peak trajectories and the peak-peak correlation discussed above, which can be readily extended to holes. The only noticeable difference between continuous trajectories of peaks and holes is that the former trace, as mentioned, the mass growth of accreting halos, while the latter do not trace the mass evolution of voids because underdense regions do not collapse; they only deepen. In other words, continuous hole trajectories only trace the same fixed voids seen at different scales.

\section{Strategy}\label{stra}

The usual approach followed to calculate the typical AM of protohalos of mass $M$ collapsing at $\tc$ (or with $\delta$ at $R$) is to compute the protohalo inertia tensor {\bf I}, the tidal (or deformation) tensor {\bf T} at the protohalo c.o.m., and use the joint PDF of all quantities appearing in those tensors, $P(q_1,q_2,q_3,...|\delta,R)$, to average the modulus of the protohalo {\bf J} given by Equation (1).

Instead, we will calculate the tidal tensor ${\bf T}\nb$ due to each individual neighboring tidal source, proceed in the usual way to find the AM caused by it, ${\bf J}_{\rm ts}$, integrate the AM due to all sources, and average the resulting global $J$ over all possible configurations of this composite system.

The joint PDF $P\nb(q_1,q_2,q_3,...|\delta,R)$ of properties $q_i$ of each single torque source will be split into the product of the conditional probability $P\nb$ of finding such properties $q_1$, $q_2$, $q_3$,... subject to having found the protohalo with the properties $\tilde q_1$, $\tilde q_2$, $\tilde q_3$,... times the probability $P\pk$ that the protohalo of given $\delta$ at $R$ has such properties,
\beqa
P\nb(q_1,q_2,...|\delta,R)=P\nb(q_1,q_2,...|\tilde q_1,\tilde q_2,...,\delta,R)\nonumber\\
\times P\pk(\tilde q_1,\tilde q_2,...|\delta,R).~~~~~~~~~~
\eeqa
This facilitates concentrating, in Section \ref{correspondence}, in the calculation of the protohalo inertia tensor and the PDF of the protohalo (or peak) properties, $P\pk(\tilde q_1,\tilde q_2,\tilde q_3,...|\delta,R)$, and, in Section \ref{potential}, in the calculation of the tidal tensor due to one torque source and the conditional PDF of the torque source properties subject to having found the protohalo with some properties, $P\nb(q_1,q_2,q_3,...|\tilde q_1,\tilde q_2,\tilde q_3,...,\delta,R)$. Finally, the resulting PDF of each torque source will be used in Section \ref{AM} to integrate and average the protohalo AM.

It is also worth mentioning that it will often happen that some property $q_i$, e.g.~$q_2$, or some property $\tilde q_i$, e.g.~$\tilde q_2$, does not correlate with the remaining properties. In this case, it will disappear from the respective conditional PDF, though not from the global joint PDF, where its own PDF will appear as an isolated factor, e.g.
\beqa
P\nb(q_1,q_2,q_3,...|\delta,R)=P\nb(q_2)P\pk(\tilde q_2)~~~~~~~~~\nonumber\\
\times P\nb(q_1,q_3,...|\tilde q_1,\tilde q_3,...,\delta,R) P\pk(\tilde q_1,\tilde q_3,...|\delta,R).
\eeqa

\section{Protohalo}\label{correspondence}

The conditional PDF of finding in an infinitesimal volume at $\ti$ a peak with some given curvature, shape (i.e.~ellipticity and prolateness), and orientation (Euler angles $\alpha$, $\beta$, and $\kappa$), globally denoted by $C=(x,e,p,\alpha,\beta,\kappa)$, subject to having density contrast $\delta$ (or height $\nu=\delta/\sigma_0$) at $R$ is $P\pk(C|\nu,R)=\!{\cal N}\pk(\nu,C,R)$, where  
\beq
\!{\cal N}\pk(\nu,C,R)
\!=\!\frac{{\rm e}^{-\frac{\nu^2}{2}}}{4\pi^2R_{\star}^3}
 \frac{F(x) {\rm e}^{-\frac{(x-\gamma\nu)^2}{2(1-\gamma^2)}}}{[2\pi(1-\gamma^2)]^{\frac{1}{2}}}
 \!P_{\rm ep}(e,p)\!P_{\rm E}(\alpha,\beta,\kappa)
\label{Npk}
\eeq
is the average number density of peaks with those properties (BBKS). In Equation (\ref{Npk}), $R_\star\equiv \sqrt{3}\,\sigma_1/\sigma_2$, $P_{\rm E}(\alpha,\beta,\kappa)$ is the usual isotropic PDF of Euler angles,\footnote{The orientation of triaxial peaks does not depend on their remaining properties, so the set of properties $C$ in the conditional probability $P\pk(C|\nu,R)$ is reduced to $C=(x,e,p)$.} and
\beq
P_{\rm ep}(e,p)\approx \frac{1}{2\pi\sigma_e\sigma_p}\,{\rm e}^{-\frac{(e-\lav e\rav)^2}{2\sigma_e^2}-\frac{(p-\lav p\rav)^2}{2\sigma_p^2}}
\label{Pep}
\eeq
is the joint PDF of ellipticities $e=(\lambda_1-\lambda_3)/(2x\sigma_2)$ ($e\ge 0$) and prolatenesses $p=(\lambda_1-2\lambda_2+\lambda_3)/(2x\sigma_2)$ ($-e\le p\le e$), where $\lambda_1\ge \lambda_2\ge \lambda_3$ are minus the eigenvalues of the Laplacian, related to the peak curvature through $\lambda_1+\lambda_2+\lambda_3=x\sigma_2$. To write Equation(\ref{Pep}), we have taken into account that, for large $x$ as corresponding to the massive halos of interest, $P_{\rm ep}(e,p)$ is nearly Gaussian with means $\lav e\rav =1/\{\sqrt{5}x[1+6/(5x^2)]^{1/2}\}$ and $\lav p\rav =6/\{5x^4[1+6/(5x^2)]^2\}$ and dispersions $\sigma_{\rm e}=\lav e\rav/\sqrt{6}$ and $\sigma_{\rm p}=\lav e\rav/\sqrt{3}$ (BBKS). This reflects the fact that, as mentioned, the peak shape correlates with the height $\nu$ only through $x$. 

As mentioned, the second factor on the right of Equation (\ref{Npk}) giving the $x$-PDF is very sharply peaked, so the average of any function $f(x)$ is essentially equal to $f(\lav x\rav)$. This allows us to marginalize the curvature and work with $x$ replaced by $\lav x\rav(R,\delta)$ everywhere.\footnote{The $e$- and $p$-PDFs are less peaked, so it is preferable not to make a similar approximation for these quantities.} 

By doing this, we are led to
\beq 
P\pk(C|\nu,R)=P_{\rm ep}(e,p){\cal N}\pk(\nu,R),
\label{peak}
\eeq 
with $C=(e,p)$ and 
\beq
\!{\cal N}\pk(\nu,R)
\!=\!\frac{G_0(\gamma,\gamma\nu)}{(2\pi)^2\,R_{\star}^3}{\rm e}^{-\frac{\nu^2}{2}}
\label{calN}
\eeq
giving the average number density of peaks with $\nu$ at $R$.

The inertia tensor {\bf I} of the ellipsoidal protohalo relative to its c.o.m. in Cartesian coordinates oriented along the principal axes is
\beq
{\bf I}=\frac{M}{5\,a^2(\ti)}\left(\begin{matrix} \iota_1 & 0 & 0\\ 0& \iota_2 &0 \\ 0 & 0 & \iota_3\end{matrix}\right),
\label{orien}
\eeq
where $\iota_1\equiv a_2^2+a_3^2$, $\iota_2\equiv a_1^2+a_3^2$ and $\iota_3\equiv a_2^2+a_1^2$, being $a_1>a_2>a_3$ the semi-axes of the ellipsoidal protohalo at $\ti$, inversely proportional to the square root of the eigenvalues $\lambda_1<\lambda_2<\lambda_3$, respectively. To write Equation (\ref{orien}) we have taken into account that Equation (1) holds to first order in perturbed quantities and the tidal tensor is necessarily of first order because caused by {\it peculiar} mass fluctuations, so we can take the density of the protohalo to leading order, i.e.~with uniform density equal to $\bar\rho(\ti)$. Thus, {\bf I} is a Lagrangian tensor (it does not depend on $t$) independent of $\ti$, as expected.

The semi-axes $a_i$ of the (non-smoothed) protohalo are completely fixed by its (accurate) mass $M$ or extension $R\F$. Indeed, the mass $M=(4\pi/3) \bar\rho(\ti)a_1a_2a_3$ of the ellipsoid is, by definition of $R\F$, equal $(4\pi/3) \bar\rho(\ti) (R\F)^3$, implying 
\beq
a_i=R\F \left(\frac{\Lambda}{\lambda_i}\right)^{1/2},
\label{ai}
\eeq
where $\Lambda\equiv (\lambda_1\lambda_2\lambda_3)^{1/3}$. Thus, defining the dimensionless semi-axes as
\beq
\hat a_i\equiv \frac{a_i}{R\F}\!\left(\frac{x\sigma_2}{\Lambda}\right)^{1/2},
\label{rel2}
\eeq
the relations given above between the peak shape and eigenvalues lead to
\beq
e=\frac{1}{2}(\hat a_1^{-2}-\hat a_3^{-2})\qquad \qquad p=\frac{1}{2}(1-{3}\hat a_2^{-2}), 
\label{rel1}
\eeq
and $1=\sum_k^3 \hat a^{-2}_k$.

We stress that protohalo ellipsoids have been delimited in the usual natural way: from their mass $M$ and their uniform density (to linear order in perturbed quantities) $\bar\rho(\ti)$, by simply taking into account their shape set in this case by the triaxial peak. In spherical objects, this leads to their top-hat radius $R\F$ through the relation $M=(4\pi/3) \bar\rho(\ti) (R\F)^3$. Similarly, in ellipsoid objects with known ellipticity and prolateness, the relation $M=(4\pi/3) \bar\rho(\ti) a_1a_2a_3$ has led to the top-hat semi-axes. Note also that the inertia tensor is $\propto (R\F)^{5}\propto M^{5/3}$ simply because the semi-axes $a_i$ are $\propto R\F$.

\section{Torque Sources}
\label{potential}

The torque suffered by a protohalo is caused by neighboring {\it positive and negative} mass fluctuations, marked by density maxima (peaks) with positive height, and density minima (holes) with negative height, respectively. For the moment, we will concentrate in the torque caused by mass excesses and postpone the case of mass defaults to the end of the Section. 

Mass excesses of scale smaller than the scale $R$ of the protohalo do not contribute to the tidal torque. Their individual effect is weak and they are very numerous and roughly isotropically distributed around the protohalo, so their added effect cancels (particularly when averaging over all configurations; see Sec.~\ref{AM}). On the other hand, many of the larger scale mass excesses actually correspond to a few real ones responsible for the torque, seen at different scales. We must thus find those `authentic' mass excesses, hereafter denoted by index `a'.\footnote{Dubbing them as `effective'  would be more appropriate, but index `e' is already occupied by ellipticities.} 

Each series of embedded large-scale mass excesses trace a continuous peak trajectory, solution of Equation (\ref{eqtraj}), i.e.~of the form $\delta'(R')=\delta'(R)\,(R'/R)^{m}$, starting at some peak, hereafter called the `reference peak', with the minimum scale $R$ and a density contrast $\delta'(R)$ different in general from the density contrast $\delta$ of the protohalo (but see below). Given that $m$ is negative (for any allowed spectral index $n$ and for CDM in the relevant mass range), $\delta'(R')$ decreases with increasing $R'$, although less rapidly than $(R')^{-3}$, so the mass excess $\delta'(R')(R')^3$ increases with increasing scale. Therefore, the scale $R\nb$ of the authentic mass excess is limited by the typical maximum scale $R\maxi(R,\delta)$ of those peak trajectories (see Sec.~\ref{traj}) and by the `top-hat' separation $r$ between their c.o.m. and the c.o.m. of the protohalo,\footnote{Ellipsoidal shells, or homoeoids, beyond $r$ do not contribute to the gravitational potential at the c.o.m. of the protohalo; see below.} so $R\nb=\min[R\maxi(R,\delta),r_{\rm R} r]$ and $\delta\nb=\delta'(R) (R\nb/R)^m$.

The conditional probability that an authentic mass excess of scale $R\nb$ with $\nu\nb$ and $C\nb=(e\nb,p\nb,\alpha\nb,\beta\nb,\kappa\nb)$ lies at a distance $r$ from the c.o.m. of the protohalo is
\beqa
P\nb(\nu\nb,C\nb,R\nb,r|\delta,C,R)~~~~~~~~~~~~~~~~~~~~~~~~~~~~~~\nonumber\\
=\!4\pi r^2\!\!\! \int_{-\infty}^\infty\!\!\!\!\der \delta'\! \left[1\!+\!\xi\pk(r;\nu\nb,R\nb,\nu,R)\right]~~~~~~~
\nonumber\\ \times {\cal N}\pk(\nu\nb,C\nb,R\nb|\delta',R) P(\delta',R,r|\delta,C,R),~~
\label{P1}
\eeqa 
where $P(\delta',R,r|\delta,C,R)$ is the probability of finding a point with $\delta'$ at a distance $r$ from the peak with $\delta$ and $C$ in the density field smoothed at scale $R$, ${\cal N}\pk(\nu\nb,C\nb,R\nb|\delta',R)$ is the average conditional number density of peaks with $\nu\nb$ and $C\nb$ at $R\nb$ subject to have $\delta'$ at $R$, and $1+\xi\pk(r;\nu\nb,R\nb,\nu,R)$ is the factor enhancing this number density due to the cross-correlation between peaks with $\delta$ at $R$ and with $\delta\nb$ at $R\nb$.\footnote{The point with $\delta'$ at $R$ essentially coincides with the reference peak of the authentic mass excess.} 

Both ${\cal N}\pk(\nu\nb,C\nb,R\nb|\delta',R)$ and $P(\delta',R,r|\delta, C,R)$ were calculated by BBKS. But we do not need the explicit form of the former. As shown in Appendix \ref{Prob}, for massive halos, as corresponding to bright galaxies, and separations $r$ not too large compared to $R$, as corresponding to the protohalo neighborhood, $P(\delta',R,r|\delta, C,R)$ turns out to be null for all values of $\delta'$ except for $\delta'\approx \delta$. In other words, it is close to a Dirac delta independent of $C$. This remarkable result is a consequence of the well-known protohalo bias (i.e.~the more massive objects are, the more clustered) together with the rapid fall of the (proto)halo mass function with increasing mass. The combination of both effects implies that close pairs of protohalos of the same scale tend to be twin, i.e.~to have the same density contrast (though not necessarily the same shape and orientation). Therefore, Equation (\ref{P1}) becomes 
\beqa
P\nb(\nu\nb,C\nb,R\nb,r|\delta,C,R)~~~~~~~~~~~~~~~~~~~~~~~~~~~~~~~~~\nonumber\\
\approx 4\pi r^2\!\left[1\!+\!\xi\pk(r;\nu\nb,R\nb,\nu,R)\right]{\cal N}\pk(\nu\nb,C\nb,R\nb|\delta,R)~~~\nonumber\\
=4\pi r^2\!\left[1\!+\!\xi\pk(r;\nu,R)\right]{\cal N}\pk(\nu, C\nb,R)\equiv P\nb(C\nb,r|\delta,R)\nonumber.\\
\label{P1bis}
\eeqa
To write the first equality on the right of Equation (\ref{P1bis}), we have taken into account that $\delta\nb$ and $R\nb$ are functions of $\delta$ and $R$ of the reference peak, through the relations $R\nb=\min[R\maxi(R,\delta),r_{\rm R} r]$ and $\delta\nb=\delta'(R)\,(R\nb/R)^m$, with $\delta'(R)=\delta$ thanks to the above mentioned Dirac delta. And, in the second equality, we have taken into account that the probability of finding the peak with $\nu\nb$ at $R\nb$ at some point is the same as finding the reference peak with $\nu$ at $R$ at the same point, so the cross-correlation $\xi\pk$ between the protohalo and authentic mass excesses equals the autocorrelation between identical peaks. Thus, making use of the linear peak bias (Section \ref{CUSP}), the conditional probability $P\nb$ becomes
\beq
P\nb(C\nb,r|\delta,R)
\approx 4\pi r^2\,
[1+B_1^2(R,\nu)\xi(r)]{\cal N}\pk(\nu, C\nb,R).
\label{P2bis0}
\eeq
Like all patches marked by peaks, authentic mass excesses are ellipsoidal with semi-axes $a_{\rm a1}$, $a_{\rm a2}$, and $a_{\rm a3}$.\footnote{This is certainly true for high peaks (BBKS); for low ones as corresponding to authentic mass excesses it is an approximation. But we are only interested in the peculiar gravitational potential they yield, which is much less sensitive to small departures of the sources from the ellipsoidal symmetry.} In addition, to first order in perturbed quantities, they have a uniform `peculiar density' equal to $\delta(\ti)\bar\rho(\ti)$ and a peculiar mass equal to $M\nb=(4\pi/3) a_{\rm a1}a_{\rm a2}a_{\rm a3}\delta\nb(\tc,\ti)\bar\rho(\ti)$. Consequently, the peculiar gravitational potential they cause, in Cartesian coordinates with origin at their c.o.m. and aligned with their own principal axes, at a point ${\bf x}=(x_1,x_2,x_3)$ external to it (the potential at internal points vanishes) is (\citealt{Ch87})
\beqa
\Phi\nb({\bf x})=\pi a_{\rm a1}a_{\rm a2}a_{\rm a3} G\delta\nb \bar\rho(\ti)~~~~~~~~~~~~~~~~~~~~~~\nonumber\\
\times\left[\int_{S({\bf x})}^\infty \frac{\der s}{\Delta(s)}-\sum_{i=1}^3 x_i^2\int_{S({\bf x})}^\infty \frac{\der s}{(a_{{\rm a}i}^2+s)\Delta(s)}\right],
\label{pot}
\eeqa
where $G$ is the gravitational constant,
\beq
\Delta(s)\equiv\left[(a_{\rm a1}^2+s)(a_{\rm a2}^2+s)(a_{\rm a3}^2+s)\right]^{1/2},
\label{Ds}
\eeq
and $S({\bf x})$ denotes the positive root of equation 
\beq
\sum_{i=1}^3 \frac{x_i^2}{a_{{\rm a}i}^2+S}=1.
\label{S}
\eeq 
Thus, given the relation $\phi({\bf x})=\Phi({\bf x})/[4\pi D(t)a^2(t)G\bar\rho(t)]$, between the linear gravitational potential $\phi({\bf x})$ in Equation (1) and the usual peculiar gravitational potential $\Phi({\bf x})$, the Hessian of $\Phi\nb({\bf x})$ at the c.o.m.~of the protohalo, {\bf x}, of modulus $r$ leads to the tidal tensor (see App.~\ref{TT})
\beq
{T\nb}_{ij}\!\approx\!\frac{\delta\nb}{3D(\ti)} \left(\frac{R\nb\F}{r}\right)^{\!\!3}\! \left\{\frac{3\Lambda\nb}{\lav x\nb\rav\sigma_{\rm 2a}}\!\left[\frac{(R\nb\F)^2}{r}\right]^2\frac{x_ix_j}{a^2_{{\rm a}i}a^2_{{\rm a}j}}
\!-\!\delta_{ij} \right\}\!.
\label{potbo}
\eeq
As can be seen, this tensor, which is proportional to $\delta\nb$ is also Lagrangian (it does not depend on $t$) and independent of $\ti$ (it is only a function of $\tc$; see Equation (\ref{deltat})).

Let us come back to mass defaults also contributing to the torque. For the reasons explained in Section \ref{holes}, the preceding derivation for mass excesses (and peaks) also holds for mass defaults (and holes), with the only difference that, in the latter case, $\delta\nb$ and $\lav x\nb\rav$ are negative. Since the number density contrast of peaks with positive $\delta$ at $R$ is $\delta\pk=B_1(R,|\nu|) \delta\m$ and that of holes with negative $\delta$ at $R$ is $\delta_{\rm h}=- B_1(R,|\nu|)\delta\m$ (see Equation (\ref{simple1})), the peak-hole cross-correlation is given by $\xi_{\rm pk\,h}(r)=-B_1^2(R,|\nu|)\xi(r)$. Consequently, the probability of finding an authentic large-scale mass excess {\it or} default, from now on simply a torque source, with $\delta\nb$ at $R\nb$ at a distance $r$ from the protohalo is $P\nb(C\nb|\delta,R,r)$ given by Equation (\ref{P2bis0}), but a factor two higher and no bias term arising from the correlation between sources because the peak-peak correlation exactly cancels with the peak-hole one, i.e.
\beq
P\nb(C\nb,r|\delta,R)
=8\pi r^2\,{\cal N}\pk(\nu, C\nb,R).
\label{P2bis}
\eeq
The fact that the protohalo-torque source correlation vanishes when both peaks and holes are taken into account is well-understood: given a peak, the probability of finding other peaks in its neighborhood is higher than in the average, but the probability of finding holes is lower, and both effects balance each other. This result has important consequences in the calculations next.

\section{Protohalo angular momentum}
\label{AM}

Given a protohalo of mass $M$ and collapse time $\tc$ (or with $\delta$ and $R$), the components, in a Cartesian reference system with origin at its c.o.m., of its Lagrangian AM with respect to that point due to the torque of one source only is (Equation (1)) 
\beq
{J\nb}_i=-\epsilon_{ijk}[A\nb T\nb{A\nb^{\rm T}}]_{jl} [{AIA^{\rm T}}]_{lk}\,,
\label{AM1}
\eeq
where {\bf I} and ${\bf T\nb}$ are the Lagrangian inertia tensor of the protohalo and the Lagrangian tidal tensor due to that torque source, both oriented along their own axes, given in Sections \ref{correspondence} and \ref{potential}, respectively, and ${\bf A}$ and ${\bf A}\nb$, with index T denoting transpose, are the rotation (or direction cosine) matrices that reorient them along the reference system. Therefore, to find the {\it typical} AM of protohalos we must sum up the contribution of every single torque source in a given configuration and average the result over all possible configurations (mutual separations, shapes, and orientations of all objects). This task may seem unfeasible, but, as shown next, the absence of correlation between the protohalo and the torque sources and between torque sources themselves makes it possible. 

The torque strength of individual sources behaves as $R\nb^3\delta\nb/(r_{\rm R}r)^3$, with $R\nb$ bounded to $\min[R\maxi(R,\delta),r_{\rm R} r]$, so, for fixed $\delta\nb$, the strength decreases with increasing $r$ or stays at most constant for torque sources at small $r$. And, since $\delta\nb=(R\nb/R)^m\,\delta$ decreases with increasing $R\nb$, the strongest torque source is necessarily the closest one to the protohalo, at a separation smaller than the typical mean separation between sources, so it satisfies $R\nb=r_{\rm R} r$ or, equivalently,$R\nb\F=r$. This result will be used to simplify the form of the tidal tensor (Equation (\ref{potbo})), but, what is more important here, it greatly simplifies the integration of the contribution of all sources in one configuration of the system and the average over all configurations. 

Given that the $N$-point correlations vanish,\footnote{The extension of the two-point to $N>2$ involves the reduced $N$-point correlations, which, in case of peaks, write approximately down as $N-1$ products of two-point correlations (e.g.~\citealt{SM94}).} all objects are uncorrelated. Therefore, they are randomly distributed around any particular subsystem, such as the one formed by the protohalo and the nearest torque source, so, when averaging over all configurations with fixed protohalo-nearest torque source subsystem, the added torque of all the remaining sources cancel. We may thus concentrate in averaging over all possible configurations of the protohalo-nearest torque source subsystem only.

The probability of finding the {\it nearest} torque source with $C\nb$ at a distance $r$ from the protohalo is the probability of finding one such torque sources inside $r$ times the probability that there is none at smaller separations (see Equation (\ref{P2bis})),
\beqa
P_{\rm st}(C\nb,r|\delta,R)= 
8\pi r^2 {\cal N}\pk(\nu,C\nb,R)~~~~~~~~~~~~~~\nonumber\\
\times \left[1-\frac{8\pi }{3} r^3 {\cal N}\pk(\nu,C\nb,R)\right]\!.
\label{prob}
\eeqa
Therefore, the joint PDF of the protohalo-nearest torque source properties to be used in the average of $J\nb$ given by Equation (\ref{AM1}) (though referring to the nearest torque source) over all configurations of this subsystem is $P_E(\alpha\nb,\beta\nb,\kappa\nb)P_E(\alpha,\beta,\kappa)$ times 
\beq
P_{\rm st}(C\nb,C,r|\delta,R)=P_{\rm st}(C\nb,r|\delta,R)P\pk(C|\delta,R),
\label{joint}
\eeq
with $C\nb=(e\nb,p\nb)$, $C=(e,p)$, and the conditional probabilities $P\pk$ and $P_{\rm st}$ given by Equations (\ref{peak}) and (\ref{prob}), respectively.

We will start by averaging $J\nb$ over the uncorrelated orientations of the protohalo and the nearest torque source, i.e.~over the Euler angles $\alpha,\beta,\kappa$ and $\alpha\nb,\beta\nb,\kappa\nb$. We must not average, of course, the components ${J\nb}_i$ themselves, but the modulus of ${\bf J}\nb$. However, by isotropy of the universe, the average of any component ${J\nb}_i$ must be the same, implying $\lav J\nb \rav^2=3\lav {J\nb}_i\rav^2$. Consequently, the average modulus can be calculated from the average of any given component. The resulting AM is (App.~\ref{orientation})
\beqa
J\nb=\sqrt{3}\lav {J\nb}_1\rav\approx \frac{0.040
}{3^{5/6}\pi^{2/3}}\frac{G \bar\rho^{1/3}_0 \delta \, M^{5/3}}{D(\ti)H_0^2\Omega_0}\nonumber \\
\times \left(\frac{r}{R\F}\right)^m H\!\left(e\nb,p\nb\right)\,(1-2p)e,
\label{j3}
\eeqa
where $\bar\rho_0$, $H_0$, and $\Omega_0$ are the current mean cosmic density, Hubble constant, and the matter density parameter, respectively, and $H(e\nb,p\nb)$ is a function of the ellipticity and prolateness of the torque source, given by Equation (\ref{H}) with $R\nb\F=r$. We remark that the factor $\delta=\delta(\tc,\ti)$ arises from the peculiar density of torque sources, written in terms of the density contrast of the protohalo, and the factor $M^{5/3}$ arises from the inertia tensor.

Then, averaging over the ellipticity and prolateness of the protohalo and of the nearest torque source, we arrive at (App.~\ref{averep})
\beq
J\nb\approx 0.018\, \frac{G \bar\rho^{1/3}_0M^{5/3}}{H_0^2\Omega_0} g(\gamma,\gamma\nu)\left(\frac{r}{R\F}\right)^{m}\!\!\frac{\delta}{D(\ti)},
\label{aver}
\eeq
where
\beq
g(\gamma,\gamma\nu)\equiv \frac{1}{\left(\lav x\rav^2+\frac{6}{5}\right)^{1/2}}\left[1-\frac{1.182}{\left(\lav x\rav^2+\frac{6}{5}\right)^2}\right]
\label{g}
\eeq
is time-invariant and weakly dependent on $M$.

Lastly, we must average over all possible separations $r$ of the nearest torque source for $P_{\rm st}(r|\delta,R)$ given by Equation (\ref{joint}), with no $C\nb$ and $C$ arguments since already averaged, i.e.
\beq
P_{\rm st}(r|\delta,R)\!=\! 
8\pi r^2 {\cal N}\pk(\nu,R)
\!\left[1\!-\!\frac{8\pi}{3} r^3 {\cal N}\pk(\nu,R)\right]\!.
\label{prob2}
\eeq
That is, we must perform the integral
\beq
J=\int_{2R\F}^{r_{\rm one}(\delta,R)} \der r J\nb(r,\delta,R) P_{\rm st}(r|\delta,R),
\label{nt}
\eeq
where we have taken into account that the minimum scale of the torque source is the scale $R$ of the protohalo so that the minimum possible value of $r$ is $2R\F$. In Equation (\ref{nt}), $r_{\rm one}$ is the radius of the sphere centered at the c.o.m. of the protohalo that harbors one main torque source, solution of the implicit equation 
\beq
1=8\pi R^3{\cal N}\pk(\nu,R)\int_{2R\F/R}^{r_{\rm one}/R}\! \der s\, s^2,
\label{rR}
\eeq
with ${\cal N}\pk(\nu,R)$ given by Equation (\ref{calN}), that is
\beq
\left(\frac{r_{\rm one}}{R\F}\right)^3=8+{r_{\rm R}^3}\left[\frac{2}{3 \pi}\!\left(\!\frac{n+5}{6}\!\right)^{3/2}\,G_0(\gamma,\gamma\nu)\,{\rm e}^{-\frac{\nu^2}{2}} \right]^{-1}.
\label{im2}
\eeq

At this point, it is worth mentioning that, following the previous derivation for protohalos conditioned to lie in a background $\delta\m$ at scale $R\m>3R$ instead of for unconditioned ones, we would have been led to Equation (\ref{rR}) with the torque source number density ${\cal N}\pk$ including an additional term proportional to $\nu\delta\m/\sigma_0$ for peaks, and another for holes (see \citealt{Sea24}). But since $\nu$ has the opposite sign in peaks and holes, those additional terms would cancel and $r_{\rm one}/R\F$ would remain the same. The reason for this is simple: in overdense backgrounds there are more peaks, but also less holes, and conversely in underdense backgrounds. The result that $r_{\rm one}$ does not vary with background density was used in \citet{Sea24} in the study of the secondary bias of halo AM.\footnote{The detailed form of function $g(\gamma,\gamma\nu)$ announced in that Paper is to be replaced by the right one derived here.}

\begin{figure}
 \includegraphics[scale=1.25,bb= 43 0 50 200]{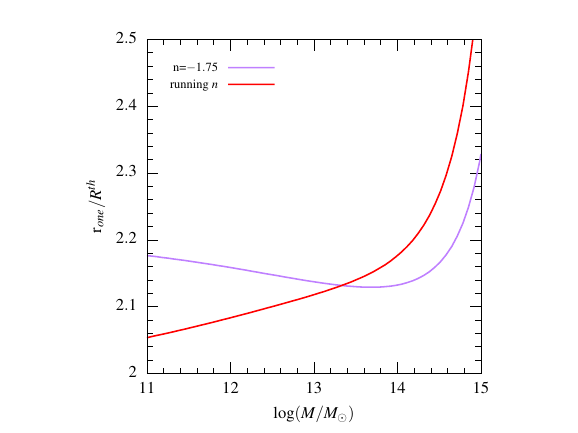}
 \caption{Maximum separation between the centers of mass of the protohalo and the main tidal torque source, in units of the protohalo (top-hat) radius, as a function of halo virial mass in the {\it Planck14} cosmology. Shown are the solutions obtained with fixed $n$ equal to $-1.75$ (violet line) and running $n(M)$, drawn from the relation $\gamma^2=(n+3)/(n+5)$ holding for power-law spectra, with the accurate spectral coefficient $\gamma(M)$ (red line).} 
 \label{rst}
\end{figure}

In Figure \ref{rst} we plot $r_{\rm one}/R\F$ in the {\it Planck14} cosmology \citep{P14} and for virial masses where $r_\sigma$ and $r_{\rm R}$ are time-independent. $r_{\rm one}/R\F$ is kept quite constant ($\sim 2.1$) except at the high-mass end, where the halo number density falls off and their typical separation rapidly increases. Similar results are obtained for other cosmologies and mass definitions. 

\begin{figure}
 \includegraphics[scale=1.25,bb= 43 0 50 200]{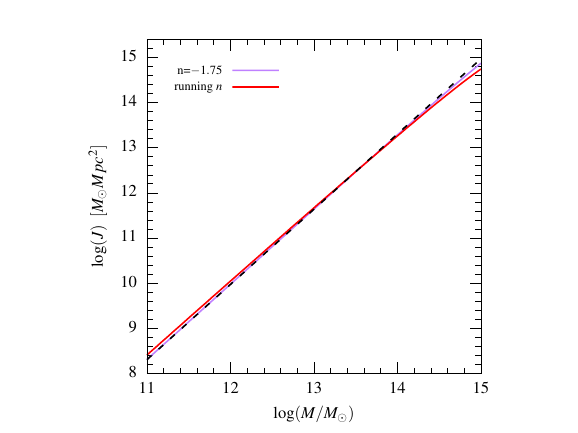}
 \caption{Mean Lagrangian protohalo AM (same lines as in Figure \ref{rst}) as a function of virial mass of current halos in the {\it Planck14} cosmology compared to the best $J\propto M^{5/3}$ fit (black dashed line).} 
 \label{Jfig}
\end{figure}

\begin{figure}
 \includegraphics[scale=1.25,bb= 43 0 50 200]{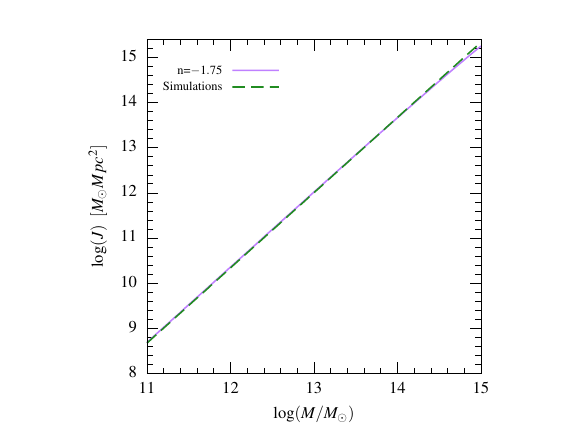}
 \caption{Comparison of the predicted mean Lagrangian protohalo AM to that found in simulations by SSK00 in the EdS cosmology with $H_0=50$ km s$^{-1}$ Mpc$^{-1}$.} 
 \label{Jfig2}
\end{figure}

The final average (\ref{nt}) leads, to leading order in $1-2R\F/r_{\rm one}\ll 1$, to the desired expression of the mean Lagrangian protohalo AM, 
\beq
J \approx 0.018\,\frac{G\bar\rho^{1/3}_0 M^{5/3}}{H_0^2 \Omega_0}g(\gamma,\gamma\nu)\left(\frac{r_{\rm one}}{R\F}\right)^{\!m}\!\frac{\delta}{D(\ti)}.
\label{Jdef}
\eeq

Note that $J$ does not depend on $\ti$ because $r_{\rm one}/R\F$ is independent of it (Equation (\ref{im2})) and $\delta$ stands for $\delta(\tc,\ti)=r_\delta(\tc)\delta\cc\F(t\cc)D(\ti)/D(\tc)$ (Equation (\ref{deltat})). 

The predicted mean Lagrangian AM (Equation (\ref{Jdef}) is shown in Figure \ref{Jfig} for the same cosmology and mass definition as in Figure \ref{rst}. As can be seen, $J$ is very nearly proportional to $M^{5/3}$ as found in simulations (e.g.~\citealt{BE87}; SSK00), regardless of whether $n$ is taken equal to the effective spectrum index or with running value dependent on mass according to the CDM spectrum. This shows that the prediction is little sensitive to the power-law approximation. Similar results are obtained for other cosmologies (with CDM and power-law spectra) and halo mass definitions.

The theoretical Lagrangian AM is compared in Figure \ref{Jfig2} to the results of simulations carried out by SSK00 in the EdS ($H_0=50$ km s$^{-1}$ Mpc$^{-1}$) cosmology.\footnote{\cite{Pea02a} do not provide enough information to carry out this comparison.} Specifically, the `empirical' Lagrangian $J(M)$ relation corresponds to the best power-law fit to the Eulerian $J(M)$ relation of simulated halos in the sample best suited to the comparison, the so-called LG $>200$ catalog, whose objects evolved without undergoing major mergers, as implicitly assumed in our theoretical model, and have more than 200 dark matter particles so that their AM is well determined, while they are still numerous enough to have good statistics. The best power index of the Eulerian $J(M)$ relation found by SSK00 for this halo sample is $5/3$ (see their table 6). The zero-point of this relation is not explicitly provided by these authors, but it can be estimated from their Figures 15 and 7. Indeed, as can be seen in Figure 15, the relations with slightly different power index fitting the AM of different halo samples basically pivot around a point at the center of the mass range\footnote{The scatter increases, indeed, from there, where it is particularly small, toward the high- and low-mass ends.} near a halo of the LG sample, marked with a triangle, lying besides another halo marked with a cross. Thus, we have adopted its AM $J\approx 10^{15}$ \modot kpc$^2$ Gyr$^{-1}$ (Figure 15) and mass $M=7\times 10^{11}$ \modotc (Figure 7) as a fair estimate of the desired zero-point. SSK00 identified halos at $t_0$ using the Friends-of-Friends algorithm with linking length $b=0.075$, which led to a limiting (local) overdensity of $\sim 1000$. Since the halo mass definition resulting from this procedure differs from the usual ones, we have converted the masses in the $J(M)$ relation derived by SSK00 to virial masses. To do this we have taken into account that, for halos approximately isothermal, the linking length used leads to a mean inner overdensity of $\Delta \sim 3400$ (see, e.g., Equation (3) of \citealt{Jea14b}), i.e. $\sim 20$ times the virial overdensity $\Delta\vir=178$. Thus, their virial radius is $\sim \sqrt{20}$ times larger than the actual radius at the limiting overdensity of 1000, implying a virial mass 4.5 times larger.\footnote{Such a mass conversion is little sensitive to the exact scale-free density profile assumed (e.g. for a logarithmic slope of $-2.5$, the factor would be $4.4$).} The $J(M)$ relation of simulated halos so obtained refers to the Eulerian AM at the present time $t_0$. But, as mentioned by SSK00, the AM of objects stays fixed after turnaround, so this $J(M)$ relation also holds at the mean turnaround time $t\ta= 1.51\pm 0.61$ Gyr of halos in that sample. Thus, dividing the Eulerian AM by the factor $a^2(t\ta)\dot D(t\ta)=(3/2)H_0^2t\ta$, we have obtained the corresponding Lagrangian relation plotted in Figure \ref{Jfig2}. On the other hand, the theoretical Lagrangian relation plotted in the same Figure has been derived using Equation (\ref{Jdef}) in the EdS cosmology with the appropriate mean collapse time $t\cc=3.01\pm 1.05$ Gyr (SSK00).

As can be seen in Figure \ref{Jfig2}, the theoretical relation overlaps with the numerical one (it is just a factor 1.05 larger). Of course, the theoretical relation is a leading order approximation and the empirical fit to the simulated data and the mass conversion applied to it are affected by a substantial error, so we must not attach too much importance to that almost full matching. But, in any event, it is clear that the predicted typical protohalo AM agrees with the AM of simulated objects. In fact, it agrees even better than found by SSK00 from direct application of Equation (1), where a factor $\sim 3$ difference was observed. Both theoretical predictions are based on the TTT, but, while applying Equation (1) requires the smoothing of the deformation tensor, Equation (\ref{Jdef}) directly uses the potential $\Phi\nb$ of mass fluctuations associated with peaks of the corresponding scales larger than $R$. This suggests that the departure of a factor $\sim 3$ between the predictions of the TTT and the results of simulations reported by SSK00 could simply arise from such an uncertain smoothing.
    
Finally, we can also calculate the median protohalo AM, $J_{\rm med}$. The Euler angles have flat PDFs and do not enter $J_{\rm med}$. The uncorrelated PDFs of the ellipticities and prolatenesses of the protohalo and the nearest torque source are Gaussian, so their median values are equal to their means. Lastly, the median separation, $r_{\rm med}$, is given by the value of $r$ for which the cumulative PDF ($\approx 8\pi r^2 {\cal N}\pk(\nu,\delta)$; Equation (\ref{prob2})) is one half, i.e.~the solution of the implicit equation
\beq
0.5\approx \frac{\int_{2R\F}^{r_{\rm med}  \der r\, r^{2}}}{\int_{2R\F}^{r_{\rm one}}  \der r\, r^{2}}=
\left(\frac{r_{\rm med}}{r_{\rm one}}\right)^{3}
\frac{1-(2R\F/r_{\rm med})^3}{1-(2R\F/r_{\rm one})^3},
\label{med}
\eeq
implying $r_{\rm med}\approx r_{\rm one}$ to leading order in $1-2R\F/r_{\rm one}$. Plugging these median values of all arguments in the non-averaged protohalo AM (Equation (\ref{j3})) and taking into account that $\lav x\nb\rav\approx (r/R\F)^m\gamma\nu$ is substantially smaller than 6/5 (like in App.~\ref{averep}), we are led to a $J_{\rm med}$ equal to $J$ at the present order of approximation.


\section{Summary and Discussion}\label{sum}

We have derived the typical Lagrangian protohalo AM predicted by the TTT within the peak model of structure formation. This has been done following a novel approach that splits the global tidal tensor into the contributions of individual neighboring mass fluctuations. The interest of this procedure is that it allows one to average the AM of protohalos of mass $M$ collapsing at $\tc$ in a fully analytic manner, which leads to a practical simple expression of this AM and facilitates the traceability of its functionality. 

Specifically, after characterizing the positive and negative mass fluctuations of all scales contributing to the global torque and taking into account the correlation between them and with the protohalo, we have integrated the AM they cause and averaged over all configurations of the system. This has led to the following simple expression of the mean (and the median as well) Lagrangian protohalo AM, 
\beq
J \approx 0.018\,\frac{G\bar\rho^{1/3}_0 M^{5/3}}{H_0^2 \Omega_0}g(\gamma,\gamma\nu)\left(\frac{r_{\rm one}}{R\F}\right)^{\!m}\!\frac{r_\delta(\tc)\delta\cc\F(\tc)}{D(\tc)}.
\label{final}
\eeq
where $\bar \rho_0$, $H_0$, and $\Omega_0$ are the present mean cosmic density, Hubble constant, and matter density parameter, respectively, $m=-(n+3)/2$, where $n$ is the real or effective power-law index of the power spectrum, $D(t)$ is the cosmic growth factor, $g(\gamma,\gamma\nu)$ is a weak function of $M$ given by Equation (\ref{g}), $r_\delta(t)\approx a(t)/D(t)$ is the ratio between the critical density contrast for ellipsoidal collapse at $t$ and its spherical counterpart $\delta\cc\F(t)\approx 1.686$, and $r_{\rm one}/R\F\approx 2$ is the separation between the centers of mass of the protohalo and the main torque source, scaled to the protohalo top-hat scale. 

But the main difference between the theoretical protohalo AM derived here from previous similar ones \citep{HP88,H88,Cea96a} is that it is based on the accurate protohalo mass and ellipsoidal collapse time provided by CUSP. In the absence of this information, ellipsoids were delimited by adopting a fixed isodensity contour, which led to protohalo semi-axes $a_i\propto \nu^{1/2}$, masses $M\propto \nu^{3/2}$, and AM $J\propto \nu^{5/2}$. In addition, the Gaussian height $\nu$ of protohalos collapsing at $\tc$ could not be related to its top-hat counterpart $\nu\F$, so the relation $M\propto \nu^{3/2}$ could not be compared to the well-known $M(\nu\F)$ relation arising from the power spectrum. Instead, using the accurate protohalo mass and collapse time of Gaussian peaks, we have obtained $\nu\approx (r_\delta/r_\sigma) \nu\F\propto \nu\F$, so $M\propto \nu^{3/2}$ is equivalent to $M\propto (\nu\F)^{3/2}$, which is inconsistent with the $M\propto (\nu\F)^{6/(n+3)}$ arising from (real or approximate) power-law spectra.

In the present derivation, ellipsoids have been delimited in the usual (consistent) way: from the mass, shape (set by the ellipticity and prolateness of the associated peaks), and density of linear protohalos. This has led to protohalo semi-axes $a_i\propto R\F$, masses $M\propto (R\F)^3$, and AM, $J\propto (R\F)^{5}$, implying that the relation $J\propto M^{5/3}$ found in simulations {\it directly arises from the inertia tensor of individual protohalos}. (In previous works, it was thought to be a statistical relation, as found by comparing the average AM of protohalos of all heights collapsing at a fixed time with the average of different powers of $M$.) 

This intrinsic origin of the $J\propto M^{5/3}$ relation contradicts the common belief that it is a consequence of the halo bias: the more massive (proto)halos are, the larger their AM because they are more clustered and suffer stronger torques. Halo clustering is responsible, indeed, for the secondary bias of their AM \citep{Sea24}, but not for the $J\propto M^{5/3}$ relation. Even though peaks are correlated, holes are anticorrelated by the same strength with respect to protohalos, so the average torque due to neighboring positive {\it and} negative mass fluctuations is the same everywhere, i.e. clustering does not affect the AM of (proto)halos.

The validity of the protohalo AM predicted here (and its physical consequences) has been checked against simulations. This is important because it is the first time that the typical protohalo AM predicted by the TTT in the peak model has been compared to simulations. Previous similar predictions were not, because, as mentioned, their format was little practical and numerical studies aimed at checking the validity of the TTT (SSK00; \citealt{Pea02a}) preferred to directly use Equation (1). The result of the comparison has been that the protohalo AM given by Equation (\ref{final}) is in good agreement with the AM of simulated objects, even better than found from Equation (1) with an uncertain smoothing of the deformation tensor. 

Since the Eulerian protohalo AM grows through the factor $a^2\dot D$ until about turnaround and then freezes out, we could use the predicted Lagrangian AM together with the accurate ellipsoidal collapse time to estimate the final AM of halos. But we can do better and take also advantage that CUSP provides the clues for properly addressing the effects of shell-crossing and halo mergers \citep{Sea12a,SM19}. This will allow us, in Paper II, to clarify how the AM freezes out in accreting halos collapsing monolithically though not homogeneously, to derive the AM of ordinary halos having undergone major mergers, and to infer the typical inner rotational properties of halos. 

\begin{acknowledgments}
We thank an anonymous referee for its invaluable job. This work was funded by the Spanish MCIN/AEI/ 10.13039/501100011033 through grants CEX2019-000918-M (Unidad de Excelencia `Mar\'ia de Maeztu', ICCUB) and PID2022-140871NB-C22 (co-funded by FEDER funds) and by the Catalan DEC through the grant 2021SGR00679.
\end{acknowledgments}

{}

\appendix 

\section{Probability of finding a density contrast near to a protohalo}\label{Prob}

The probability function of finding a point with $\delta'$ at a distance $r$ from a peak with $\delta$ and $C$ in the density field Gaussian-smoothed on the scale $R$ of the peak is normal with mean (BBKS) 
\beq
\lav\delta(r)| C\rav\!=\!\frac{\gamma\delta}{1\!-\!\gamma^2}\left(\frac{\psi}{\gamma}\!+\!\frac{\sigma_0}{\sigma_2} \nabla^2\psi\right)\!-\!\frac{\lav x\rav\sigma_0}{1\!-\!\gamma^2}\left(\gamma\psi\!+\!\frac{\sigma_0}{\sigma_2}\nabla^2\psi\right)
\!+\!\frac{5}{2}\lav x\rav\sigma_0\left(\frac{\sigma_0}{\sigma_2}\right)\left(3\frac{\psi'}{r}\!-\!\nabla^2\psi\right)P_{\rm ep}(e,p)
\eeq
(the curvature variance is unity) and variance 
\beq
\lav[\Delta\delta(r)]^2| C\rav\!=\!\sigma_0^2\bigg\{\!1\!-\!\frac{1}{1\!-\!\gamma^2}\!\left[\psi^2\!+\!\left(\!2\gamma\psi\!+\!\frac{\sigma_0}{\sigma_2} \nabla^2\psi\right)\!\frac{\sigma_0}{\sigma_2} \nabla^2\psi\right]
\!-5\!\left(\frac{\sigma_0}{\sigma_2}\right)^2\left(\!3\frac{\psi'}{r}\!-\!\nabla^2\psi\!\right)^{\!\!2}\!-\!\frac{\sigma_0}{\sigma_2}\frac{3(\psi')^2}{\gamma}\!\bigg\},
\eeq
where $\psi$ is the matter correlation function $\xi(r)$ at scale $R$ normalized to $\xi(0)$ and $\psi'$ is its $r$-derivative (in BBKS, the separation $r$ is in units of $R_\star$). As discussed in BBKS, for high peaks (massive halos), $\lav x\rav$ approaches $\gamma\nu$, and the gradients of $\psi$ can be neglected in front of unity and of $\psi$, which is of order unity when $r$ is small (of order $R$). Consequently, $\lav [\Delta\delta(r)]^2| C]\rav$ approaches $\hat\sigma_0\equiv \sigma_0^2[1-\psi^2]\approx 0$ and $\lav \delta(r)|C\rav$ essentially becomes $\delta\psi\approx \delta$. 

Therefore, the conditional probability $P(\delta',R,r|\delta,C,R)$ in Equation (\ref{P1}) approaches 
\beq
P(\delta',R,r|\delta,C,R)\approx \left[\frac{{\rm e}^{-\frac{(\delta'-\delta)^2}{2\hat\sigma_0^2}}}{\sqrt{2\pi}\hat\sigma_0}\right]_{\hat \sigma_0\approx 0}\!\!\!
\approx \delta_{\rm D}(\delta'-\delta),
\label{P}
\eeq
where $\delta_{\rm D}(\delta'-\delta)$ is the Dirac delta. Note that the properties $C$ and the separation $r$ do not appear now in the final approximate expression because they do not correlate with other properties, so $P(\delta',R,r|\delta, C,R)$ becomes $P(\delta',R|\delta,R)$. 

\section{Tidal Tensor}\label{TT} 

Given that the linear gravitational potential $\phi$ entering Equation (1) is $1/[4\pi D(t)a^2(t)G\bar\rho(t)]$ times the usual peculiar gravitational potential $\Phi$ (Equation (\ref{pot})), the Hessian of $\Phi$ at a point {\bf x} from the torque source leads to the following tidal tensor 
\beq
{T\nb}_{ij}\!=\!\frac{\delta\nb}{D(\ti)}\frac{\Delta(0)}{\Delta[S({\bf x})]}
\!\!\left(\!\frac{x_ix_j}{[a_{{\rm a}i}^2\!+\!S({\bf x})][a_{{\rm a}j}^2\!+\!S({\bf x})]}
\!\left\{\sum_{k=1}^3 \!\frac{x_k^2}{[a_{{\rm a}k}^2+S({\bf x})]^2}\!\right\}^{\!\!-1}
\!\!\!\!-\delta_{ij}\!\int_{S({\bf x})}^\infty \!\!\frac{\Delta[S({\bf x})]\, \der s}{\Delta(s)(a^2_{{\rm a}i}\!+\!s)}\!\right)\!,
\label{pot2}
\eeq
where $\delta_{ij}$ is the Kronecker delta, $\Delta[S({\bf x})]$ is given by Equation (\ref{Ds}) but with $s$ replaced by $S({\bf x})$ defined in Equation (\ref{S}).

The relation $a_{\rm a1}a_{\rm a2}a_{\rm a3}=(R\nb\F)^3$ implies that the semi-axes of the putative ellipsoidal isodensity contours at {\bf x} with modulus $r$ are $r/R\nb\F$ times the real semi-axes of the torque source. This implies in turn that $\sum_{k=1}^3 x_k^2/a^4_{{\rm a}k}=(r/R\nb\F)^2\sum_{k=1}^3 1/a^2_{{\rm a}k}=[r/(R\nb\F)^2]^2\lav x\nb\rav \sigma_{\rm 2a}/\Lambda\nb$, and $\Delta(0)=(R\nb\F)^3$. On the other hand, taking into account that the ellipticity and prolateness of peaks are moderate (BBKS) so that $S/a^2_{{\rm a}i}\approx S/(R\nb\F)^2$, we have $S({\bf x})\approx r^2-(R\nb\F)^2$, $\Delta[S({\bf x})]=r^3$, and $\Delta(0)=(R\nb\F)^3$. Consequently, the tidal tensor at the c.o.m. of the protohalo takes the form 
\beq
{T\nb}_{ij}\approx \frac{\delta\nb}{D(\ti)} \left(\frac{R\nb\F}{r}\right)^3
\left\{\frac{\Lambda\nb}{\lav x\nb\rav\sigma_{\rm 2a}}\left[\frac{(R\F\nb){^2}}{r}\right]^2\frac{x_ix_j}{a^2_{{\rm a}i}a^2_{{\rm a}j}}
-\delta_{ij}U(a^2_{{\rm a}j})\right\}\!,
\label{pot4}
\eeq
In Equation (\ref{pot4}),
\beq
U(a^2_{{\rm a}j})\equiv\frac{1}{2}\int_0^\infty\frac{\der s/a^2_{{\rm a}j}}{1+s/a_{{\rm a}j}^2}
\left[\prod_{k=1}^3\left(1+\frac{s}{a^2_{{\rm a}k}}\right)\right]^{-1/2}
=\frac{1}{2}\int_1^\infty\,\frac{\der y}{y^{3/2}}\left[\sum_{k=1}^3 \frac{a_{{\rm a}j}^2+s(y)}{a_{{\rm a}k}^2+s(y)}\right]^{-1},
\label{U}
\eeq
which, given that the ellipticity and prolateness of large-scale peaks is moderate (BBKS), becomes 
\beq
U(a^2_{{\rm a}j})\approx\frac{1}{6}\int_1^\infty\,\frac{\der y}{y^{3/2}}=\frac{1}{3}.
\eeq


\section{Average over Orientations}\label{orientation}


Since for any spatial configuration of the protohalo-torque source subsystem there is another one yielding an AM with the same modulus and opposite sign, the average of $|{J\nb}_i|$ must be carried out over half the whole composite solid angle $(8\pi^2)^2$ sd. 

Since the averages over the two sets of Euler angles may be carried out independently, we can start by averaging over $\alpha$, $\beta$ and $\kappa$ in the whole solid angle $8\pi^2$ sr. The scaled inertia tensor (Equation (\ref{orien})) then becomes
\beq
\lav {I}\rav_{\alpha,\beta,\kappa}=\frac{M}{15a^2(\ti)}\left(\begin{matrix} \iota_1 & 0 & 0\\ 0& \iota_2 &0 \\ 0 & 0 & \iota_3\end{matrix}\right),
\label{orien2}
\eeq
and the component $i$ of the AM reads
\beq
\lav J_{{\rm a}i}\rav_{\alpha,\beta,\kappa}= -\frac{M}{15a^2(\ti)} \epsilon_{ijk} \,[{A\nb T\nb A\nb^{\rm T}}]_{jk} \iota_k
=-\frac{M}{15a^2(\ti)}[{A\nb T\nb A\nb^{\rm T}}]_{jk} \left(\iota_j-\iota_k\right),
\label{j31}
\eeq
with $i$, $j$, and $k$ in the usual ciclic order. 

Since the principal axes of {\bf I} may not coincide with those of ${\bf T\nb}$, regardless of the orientation of ${\bf T\nb}$, we may assume $\iota_j-\iota_k\equiv a_k^2-a_j^2$ in Equation (\ref{j31}) equal to $a_3^2-a_1^2$. This is very convenient because then we simply have (see Equations (\ref{rel2})-(\ref{rel1}))
\beq
a_3^2-a_1^2=\frac{1}{a_3^{-2}}-\frac{1}{a_1^{-2}}=\frac{a_1^{-2}\!-\!a_3^{-2}}{a_3^{-2}a_1^{-2}}
\approx\frac{(R\F)^2}{3}\frac{\hat a_1^{-2}\!-\!\hat a_3^{-2}}{\hat a_3^{-2}\hat a_1^{-2}}\approx 6(R\F)^2(1-2p)e
\eeq
to first order in $e$ and $|p|$ (for peaks of galactic mass, $0< e\la 0.3$ and $|p|\la 0.1$), in which case $\Lambda/(\lav x\rav\sigma_2)\approx 1/3$. 

Consequently, we are led to 
\beq
\lav J_{{\rm a}i}\rav_{\alpha,\beta,\kappa}\approx \frac{2}{5}\frac{(R\F)^2M}{a^2(\ti)} [{A\nb T\nb A\nb^{\rm T}}]_{jk}(1-2p)e \,\,=\frac{2}{5}\left(\!\frac{3}{4\pi}\!\right)^{\!2/3}\frac{\bar\rho_0^{1/3}M^{5/3}}{\bar\rho_0} [{A\nb T\nb A\nb^{\rm T}}]_{23}(1-2p)e
\label{j31bis}
\eeq
in terms of the ellipticity and prolateness of the peak associated with the protohalo.

We must now average over the Euler angles $\alpha\nb$, $\beta\nb$ and $\kappa\nb$ in $8\pi^2/2$ sr from any arbitrary initial orientation of the position vector ${\bf r}\nb$ of the c.o.m. of the protohalo (e.g.~in the $a_{{\rm a}i}$ direction). After a lengthy calculation and taking into account Equation (\ref{rel2}), we arrive at the following average of $[{ A\nb T\nb A\nb^{\rm T}}]_{23}$, with ${\bf T\nb}$ given by Equation (\ref{potbo}),
\beq
\lav [{A\nb T\nb A\nb^{\rm T}}]_{23}\rav_{\alpha\nb,\beta\nb,\kappa\nb}= \frac{1}{2^5 3\pi}\frac{\delta}{D(\ti)} \left(\frac{R\nb\F}{r}\right)^3 \left(\frac{R\nb}{R}\right)^m  H\!\left(a_{{\rm a}i}\right) 
\eeq
\beq
H(a_{{\rm a}i})\!\equiv\!\frac{4\pi\!-\!9}{36}+3\frac{\lav x\nb\rav\sigma_{\rm 2a}}{\Lambda\nb}\!\left(\!\frac{R\nb\F}{r}\!\right)^{\!2}\!\left[\!\frac{5\pi-8}{30}\hat a\no^{-4}\!+\!\left(\!\frac{8}{9\pi}\!+\!\frac{26\!+\!75\pi}{120}\!\right)\!\hat a\no^{-2}\hat a\nd^{-2}\!-\!\frac{1}{10}\hat a\nd^{-4}\!-\!\frac{1}{5}\hat  a\no^{-2}\hat a\nt^{-2}\!-\!\left(\!\frac{1}{5\pi}\!+\!\frac{3\pi}{16}\!\right)\!\hat a\nd^{-2}\hat a\nt^{-2}\!+\!\frac{4}{5}\hat a\nt^{-4}\!\right]\!,
\label{H1}
\eeq
where we have taken into account $\delta\nb=(R\nb/R)^m\delta$. Given the equalities $\hat a_1^{-2}=(1+p+3e)/3$, $\hat a_2^{-2}=(1-2p)/3$, and  $\hat a_3^{-2}=(1+p-3e)/3$, we can write $H(a_{{\rm a}i})$ in terms of $e\nb$ and $|p\nb|$. To first order in these quantities so that $\Lambda\nb/(\lav x\nb\rav\sigma_{\rm 2a})\approx 1/3$, we obtain
\beq
H(e\nb,p\nb)\approx 0.0991+ 2.567\left(\frac{R\nb\F}{r}\right)^2(1+0.120p\nb+2.375e\nb).
\label{H}
\eeq

Having performed these averages, the (positive) average of the first Cartesian component of the Lagrangian protohalo AM takes the form 
\beq
\lav {J\nb}_1\rav \equiv \lav\lav {J\nb}_1\rav_{\alpha,\beta,\kappa}\rav_{\alpha\nb,\beta\nb,\kappa\nb}
\approx \frac{0.040}{3^{4/3}\pi^{2/3}}\frac{G \bar\rho^{1/3}_0 \delta \,M^{5/3}}{D(\ti)H_0^2\Omega_0} \left(\frac{R\nb\F}{r}\right)^3 \left(\frac{R\nb}{R}\right)^m H\!\left(e\nb,p\nb\right) (1-2p)e,
\label{j32}
\eeq
where $H_0$ and $\Omega_0$ are the current Hubble constant and the matter density parameter, respectively, implying an average modulus of the Lagrangian AM equal to 
\beq
J\nb=\sqrt{3}\lav {J\nb}_1\rav\approx \frac{0.040}{3^{5/6}\pi^{2/3}}\frac{G \bar\rho^{1/3}_0 \delta \, M^{5/3}}{D(\ti)H_0^2\Omega_0}\left(\frac{R\nb\F}{r}\right)^3 \left(\frac{R\nb}{R}\right)^m \,H\!\left(e\nb,p\nb\right)\,(1-2p)e.
\label{j34}
\eeq
And, in the case of the nearest torque source, we must take $R\nb\F=r$ (Section \ref{AM}).

\section{Average over Ellipticities and Prolatenesses}\label{averep}

Given the uncorrelated ellipticity and prolateness PDFs (Equation (\ref{Pep})), the averages of $J\nb$ given by Equation (\ref{j34}) over the ellipticity and prolateness of the prothalo and the nearest torque source lead to 
\beq
J\nb\approx \frac{0.107}{3^{5/6}\pi^{2/3}} \frac{G \bar\rho^{1/3}_0 \delta\, M^{5/3}}{D(\ti)H_0^2\Omega_0} \left(\frac{R\nb}{R}\right)^m (1+0.116\lav p\nb\rav+2.287\lav e\nb\rav-2\lav p\rav) \lav e\rav.
\eeq
Lastly, expressing $\lav e\rav$, $\lav e\nb\rav$, and $\lav p\nb\rav$ as functions of $\lav x\rav$ and $\lav x\nb\rav$ and, taking into account that, for large masses as it is the case particularly for torque sources, $\lav x\nb\rav$ is approximately equal to $\gamma\nu\nb=(R\nb/R)^{m}\gamma\nu= (\min[R\maxi(R,\delta),r_{\rm R} r]/R)^m\gamma\nu$, where $\min[R\maxi(R,\delta),r_{\rm R} r]/R$ is larger than unity, so that $\lav x\nb\rav^2$ can be neglected in front of $6/5$, we obtain 
\beq
J\nb\approx 0.018\,\frac{G \bar\rho^{1/3}_0 M^{5/3}}{H^2_0\Omega_0} \left(\frac{R\nb}{R}\right)^{m} \left\{\frac{1}{\left(\lav x\rav^2+\frac{6}{5}\right)^{1/2}}\left[1-\frac{1.182}{\left(\lav x\rav^2+\frac{6}{5}\right)^2}\right] \right\}\frac{\delta}{D(\ti)}.
\label{fin}
\eeq

\end{document}